\documentclass[twocolumn,aps,prl,superscriptaddress,nofootinbib,showpacs]{revtex4-1}
\usepackage{graphicx}
\usepackage{color}
\usepackage{amsmath}
\usepackage{bm}
\usepackage{slashed}
\usepackage{epsfig}
\usepackage{amsfonts}
\usepackage{epstopdf}
\usepackage{hyperref}

\newcommand{\sect}[1]{\noindent{\it \textbf{#1} ---}}

\begin{document}
\title{Exploring hadrons' partonic structure using ab initio lattice QCD calculations}

\author{Yan-Qing Ma}
\email{yqma@pku.edu.cn}
\affiliation{School of Physics and State Key Laboratory of Nuclear Physics and
Technology, Peking University, Beijing 100871, China}
\affiliation{Center for High Energy physics, Peking University, Beijing 100871, China}
\affiliation{Collaborative Innovation Center of Quantum Matter,
Beijing 100871, China}
\author{Jian-Wei Qiu}
\email{jqiu@jlab.org}
\affiliation{Theory Center, Jefferson Lab, 12000 Jefferson Avenue, Newport News, VA 23606, USA}

\date{\today}

\begin{abstract}
Following our previous proposal \cite{Ma:2014jla}, we construct a class of good ``lattice cross sections" (LCSs), from which we could study partonic structure of hadrons from ab initio lattice QCD calculations. These good LCSs, on the one hand, can be calculated directly in lattice QCD, and on the other hand, can be factorized into parton distribution functions (PDFs) with calculable coefficients, in the same way as QCD factorization for factorizable hadronic cross sections.  PDFs could be extracted from QCD global analysis of the lattice QCD generated data of LCSs.  We also show that proposed functions for lattice QCD calculation of PDFs in the literature are special cases of these good LCSs.
\end{abstract}

\maketitle
\allowdisplaybreaks

\sect{Introduction}	
Parton distributions functions (PDFs), interpreted as probability distributions to find an active parton from a colliding hadron to carry $x$-fraction of the parent hadron momentum, are very important nonperturbative quantities. They connect hadronic cross sections with a large momentum transfer to perturbatively calculable partonic dynamics, so that we can interpret high energy scattering data and make predictions for future observables \cite{Brambilla:2014jmp,Collins:1989gx}. PDFs have been extracted by performing global analysis of high energy scattering data in the framework of QCD factorization \cite{Dulat:2015mca,Martin:2009iq,Ball:2014uwa,Alekhin:2013nda,Ethier:2017zbq}. Since PDFs are well defined in QCD, it is not only very natural, but also critically important to ask and verify if PDFs could be derived from the first principle calculations in lattice QCD (LQCD).  However, a direct calculation of PDFs in LQCD is challenging due to the fact that PDFs are defined in terms of operators with a Minkowski time dependence, while LQCD calculations are done with a Euclidean time.

Moments of PDFs, given by matrix elements of {\it local} operators, can be in principle calculated in LQCD. However, in practise, calculations are limited to the lowest three moments \cite{Nocera:2017war} because of power-divergent mixing between twist-2 operators.  In Ref.~\cite{Ji:2013dva}, Ji introduced a set of quasi-PDFs, which are defined with the same operators defining PDFs except the active parton fields are not located on the light-cone, but on a spatial axis (along the $z$ or ``$3$"-direction) with no time separation, and could be calculated in LQCD \cite{Lin:2014zya,Alexandrou:2015rja,Chen:2016utp,Alexandrou:2016jqi,Zhang:2017bzy}. It was also suggested that quasi-PDFs could approach to corresponding PDFs when the hadron momentum $P_3$ goes to infinity  \cite{Ji:2013dva,Ji:2014gla,Ji:2017rah}. In Ref.~\cite{Ma:2014jla}, it was demonstrated that if quasi-PDFs are multiplicatively renormalizable, they could be related to PDFs by an all-order QCD factorization at a finite $P_3$. Major progress has been made in understanding the complexity of ultraviolet (UV) divergences of quasi-PDFs \cite{Ishikawa:2016znu,Chen:2016fxx,Monahan:2016bvm,Briceno:2017cpo, Xiong:2017jtn, Constantinou:2017sej, Alexandrou:2017huk, Chen:2017mzz,Ji:2017oey,Ishikawa:2017faj,Green:2017xeu}. Meanwhile, various different methods have been introduced to study hadron structures using LQCD calculations, including the pseudo-PDFs approach \cite{Orginos:2017kos} and  the ``OPE without OPE" approach  \cite{Chambers:2017dov} for calculating hadron structure functions \cite{Liu:1993cv,Liu:1999ak,Liu:2016djw}. For light-cone distribution amplitudes, which are much simpler than PDFs, a coordinate-space method was also employed \cite{Aglietti:1998ur,Abada:2001if,Braun:2007wv}.

In Ref.~\cite{Ma:2014jla}, we proposed a factorization-based program to {\it extract} PDFs and other parton correlation functions by using QCD global analysis of ``data'' generated by LQCD calculations of ``lattice cross sections" (LCSs), which are defined as factorizable and ``time-independent'' hadronic matrix elements (defined by equal-time operators or with the time properly integrated).  More precisely, a good LCS for extracting PDFs should have the following properties \cite{Ma:2014jla}:
\begin{itemize}
\item
is calculable in LQCD with an Euclidean time,
\item
has a well-defined continuum limit as the lattice spacing $a\to 0$, and
\item
has the same and factorizable logarithmic collinear (CO) divergences as PDFs.
\end{itemize}
It is the last property that enables us to relate good LCSs to PDFs, just like how
hadronic cross sections are related to PDFs in terms of QCD factorization. We also argued that quasi-quark distribution is an example of good LCSs.

In this paper, we further strengthen our proposal by constructing a class of {\it good} LCSs, with which we could build up a comprehensive program to explore the partonic structure of various hadrons with many LQCD calculable observables.  We demonstrate that these LCSs have the three required properties listed above for being {\it good} LCSs.  We also comment that quasi-PDFs proposed in Ref.~\cite{Ji:2013dva}, pseudo-PDFs used in Ref.~ \cite{Orginos:2017kos} and the matrix element used in Ref.~\cite{Chambers:2017dov} are special cases of these good LCSs. The proposed method of using good LCSs could be the most general way to extract PDFs from LQCD calculations.

\sect{Hadronic matrix elements in coordinate-space}
We consider single-hadron matrix elements of renormalized nonlocal operators ${\cal O}_n({\xi})$,
\begin{align}\label{eq:lcs}
{\sigma}_{n}(\omega,\xi^2,P^2)=\langle P| {T}\{{\cal O}_n({\xi})\}|P\rangle,
\end{align}
where the subscript $n$ is a label for different operators, $T$ stands for time-ordering, $P$ is the hadron momentum, $\xi$ with $\xi^2\neq0$ is the largest separation of all fields in the operator ${\cal O}_n$, the Lorentz scalar $\omega\equiv P\cdot\xi$, and renormalization scale for ${\cal O}_n({\xi})$ is suppressed.

One choice for ${\cal O}_n({\xi})$ is the dimension-2 operators for correlations of two currents with a separation $\xi$,
\begin{align}\label{eq:2currents}
{\cal O}_{j_1j_2}(\xi)\equiv
&\,
\xi^{d_{j_1}+d_{j_2}-2}\, Z_{j_1}\, Z_{j_2} j_1(\xi)\, j_2(0)\, ,
\end{align}
where $d_j$ and $Z_j$ are the dimension and renormalization constant of the current $j$, respectively, and the overall dimensional factor is introduced so that the matrix elements in Eq.~\eqref{eq:lcs} is dimensionless with our normalization, $\langle P|P'\rangle = (2E_P)(2\pi)^3\delta^3(P-P')$.  With the scalar and vector currents, for example, we could have,
\begin{subequations}\label{eq:operators}
\begin{align}
{\cal O}_{S}(\xi)=&\,\xi^4 Z^{2}_S[\overline{\psi}_q{\psi}_{q}](\xi)\,[\overline{\psi}_{q}{\psi}_q](0)\, ,\\
{\cal O}_{V}(\xi)=&\,\xi^2Z^{2}_V[\overline{\psi}_q\slashed{\xi}{\psi}_{q}](\xi)\,[\overline{\psi}_{q}\slashed{\xi}{\psi}_q](0)\, ,\\
{\cal O}_{\widetilde V}(\xi)=&\,-\frac{\xi^4}{2}Z^{2}_V[\overline{\psi}_q\gamma_\nu{\psi}_{q}](\xi)\,[\overline{\psi}_{q}\gamma^\nu{\psi}_q](0)\, ,\\
{\cal O}_{V^\prime}(\xi)=&\,\xi^2Z^{2}_{V^\prime}[\overline{\psi}_q\slashed{\xi}{\psi}_{q^\prime}](\xi)\,[\overline{\psi}_{q^\prime}\slashed{\xi}{\psi}_q](0)\, , \, \dots \, ,
\end{align}
\end{subequations}
where $\xi^4\equiv (\xi^2)^2$, $q=u, d, s,\cdots$ stands for a quark with a definite flavor and $q^\prime$ for a quark with a different flavor, the subscripts, $S$, $V$ and $V^\prime$ refers to scalar, vector and flavor-changing vector currents, respectively, and ``$\dots$'' indicates for other possible combinations of two currents including the gluonic current, e.g., $j_{\mu\nu}\propto F_{\mu\rho}{F^{\rho}}_\nu$. Matrix elements constructed from operators in Eq.~\eqref{eq:operators} satisfy the relation
\begin{align}\label{eq:rel1}
{\sigma}^*_{n}(\omega,\xi^2,P^2)={\sigma}_{n}(-\omega,\xi^2,P^2).
\end{align}

Instead of the correlation of two currents, the nonlocal operator in Eq.~(\ref{eq:lcs}) could also be made of the correlation of gauge dependent field operators with proper gauge link(s), e.g.,
\begin{align}\label{eq:operators2}
{\cal O}_{q}(\xi)=&\,Z_q(\xi^2)\overline{\psi}_q(\xi)\,\slashed{\xi} \Phi(\xi,0)\,{\psi}_q(0)\, ,
\end{align}
where $\Phi(\xi,0)={\cal P}e^{-ig\int_0^{1} \xi\cdot A(\lambda\xi)\,d\lambda}$ is the path ordered gauge link, $Z_q(\xi^2)$ is the renormalization constant of this operator, depending on $\xi^2$~\cite{Ishikawa:2017faj}, and matrix element constructed from which satisfies the relation
\begin{align}\label{eq:rel2}
{\sigma}^*_{n}(\omega,\xi^2,P^2)=-{\sigma}_{n}(-\omega,\xi^2,P^2).
\end{align}

Besides scalar operators constructed above, we can also construct vector or tensor operators, e.g.,
\begin{align}\label{eq:operatorsT}
{\cal O}_{\mu\nu}(\xi)=&\,\xi^4Z^{2}_V[\overline{\psi}_q\gamma_\mu{\psi}_{q}](\xi)\,[\overline{\psi}_{q}\gamma_\nu{\psi}_q](0)\,.
\end{align}
To simply the discussion, we will consider only scalar operators in the following, although tensor operators can be studied following the same way.

\sect{Factorization}
We show that ${\sigma}_{n}$ defined in Eq.~\eqref{eq:lcs} could be factorized into PDFs with perturbatively calculable coefficients so long as $\xi^2$ is sufficiently small,
 \begin{align}\label{eq:fac}
\begin{split}
{\sigma}_{n}(\omega,\xi^2,&P^2)=\sum_{a}\int_{-1}^1 \frac{dx}{x}\, f_{a}(x,\mu^2) \\
&\times K_{{n}}^{a}(x\omega,\xi^2,x^2P^2,\mu^2) +O(\xi^2\Lambda_{\text{QCD}}^2)\, ,
 \end{split}
\end{align}
where $\mu$ is the factorization scale, $K_n^{a}$ are perturbatively calculable hard coefficients, and $f_{a}$ is PDF of flavor $a=q,g$ with anti-quark PDFs expressed by quark PDFs using the relation $f_{\bar{a}}(x,\mu^2)=-f_{{a}}(-x,\mu^2)$.

Let $\xi^2$ be small but not vanishing, and applying operator product expansion (OPE) to the nonlocal operator ${\cal O}_n(\xi)$ in Eq.~(\ref{eq:lcs}) \cite{Collins:1984xc}, we have
\begin{align}\label{eq:ope}
{\sigma}_{n}(\omega,\xi^2,P^2)=&\sum_{J=0}\sum_a W_{n}^{(J,a)}(\xi^2,\mu^2)\, \xi^{\nu_1}\cdots\xi^{\nu_J}\nonumber\\
&\times\langle P|{\cal O}^{(J,a)}_{\nu_1\cdots\nu_J}(\mu^2)|P\rangle\, ,
\end{align}
where $\mu$ is the renormalization scale.  The ${\cal O}^{(J,a)}_{\nu_1\cdots\nu_J}(\mu^2)$ is a local, symmetric and traceless operator of spin $J$ with ``$a$'' labeling different operators of the same spin, and
\begin{align}\label{eq:reduceME}
\langle P|{\cal O}^{(J,a)}_{\nu_1\cdots\nu_J}(\mu^2)|P\rangle&=2A^{(J,a)}(\mu^2)\nonumber \\
&\times(P_{\nu_1}\cdots P_{\nu_J}-\text{traces})\, ,
\end{align}
where the scalar quantity $A^{(J,a)}(\mu^2)=\langle P|{\cal O}^{(J,a)}(\mu^2)|P\rangle$ is the reduced matrix element.
Substituting Eq.~(\ref{eq:reduceME}) into Eq.~(\ref{eq:ope}), we have
\begin{align}\label{eq:xsec}
{\sigma}_{n}(\omega,\xi^2,P^2)=&\sum_{J=0} \sum_aW_{n}^{(J,a)}(\xi^2,\mu^2)\,
2A^{(J,a)}(\mu^2)\nonumber\\
&\times \Sigma_J(\omega,P^2\xi^2)\, ,
\end{align}
where
\begin{align}\label{eq:coeJ}
\Sigma_J(\omega,P^2\xi^2)&\equiv \xi^{\nu_1}\cdots\xi^{\nu_J} (P_{\nu_1}\cdots P_{\nu_J}-\text{traces})\nonumber\\
&=\sum_{i=0}^{[J/2]} C_{J-i}^i (\omega)^{J-2i} \left(-P^2\xi^2/4\right)^i\, ,
\end{align}
where $C$ is the binomial function and $[J/2]$ is the greatest integer less than or equal to $J/2$. Up to now, no approximation has been made in deriving Eq.~(\ref{eq:xsec}).

Since higher dimensional matrix element is relatively smaller by powers of $\Lambda_{\text{QCD}}^2\xi^2$ when two reduced matrix elements are compared, for the following discussion, we ignore this power suppressed correction to keep only terms with the lowest dimensional operators, which corresponds to keep the twist-2 operators in QCD \cite{Collins:1984xc}. Reduced matrix elements of these twist-2 operators can be expressed as moments of PDFs,
\begin{align}\label{eq:moment}
A^{(J,a)}(\mu^2)=\frac{1}{S_a}\int_{-1}^1 d x x^{J-1} f_{a}(x,\mu^2)\, ,
\end{align}
where symmetry factor $S_a=1,2$ for $a=q,g$, respectively, and $J\geq1$ because there is no scalar twist-2 operator. By substituting Eq.~(\ref{eq:moment}) into Eq.~\eqref{eq:xsec}, and comparing it with Eq.~(\ref{eq:fac}) with the $K_n^a$ expanded in power series of $\omega$, we prove that $\sigma_n$ in Eq.~\eqref{eq:xsec} has the factorized form in Eq.~\eqref{eq:fac} with
\begin{align}\label{eq:coeK}
K_{n}^a=\sum_{J=1} \frac{2}{S_a} W_{n}^{(J,a)}(\xi^2,\mu^2)\, \Sigma_J(x \omega,x^2P^2 \xi^2)\, .
\end{align}
Note, however, that our proof is valid only when $|\omega|\ll1$ and $|P^2\xi^2|\ll1$ because OPE works only in the region where all components of $\xi$ go to zero uniformly but with all other variables fixed.
That is, we need to further extend our proof to other regions, especially when $\omega$ is finite.

The validity of OPE guarantees that ${\sigma}_{n}$ is an analytic function of $\omega$ in the neighborhood of $\omega=0$, and its Taylor series of $\omega$ around $\omega=0$ is defined by Eqs.~(\ref{eq:ope}-\ref{eq:coeK}). If we fix $\xi$ to be at short-distance while we increase $\omega$ by adjusting external momentum $P$, we cannot introduce any new perturbative divergence.  That is, ${\sigma}_{n}$ remains to be analytic as $\omega$ becomes larger, and only possible singularity is at $\omega=\infty$. Similarly, for fixed $\xi$, ${\sigma}_{n}$ is an analytic function of $P^2\xi^2$ except for the point of infinity. Therefore, the factorization in Eq.~\eqref{eq:fac}, defined by a Taylor series of $\omega$ and $P^2\xi^2$, holds for any finite value of $\omega$ and $P^2\xi^2$ with the correction up to $O(\xi^2\Lambda_{\text{QCD}}^2)$.

Note that the analytic behavior of $\sigma_n$ discussed above could be significantly different when it is Fourier transformed into momentum space, i.e.,
\begin{align}\label{eq:lcsM}
\widetilde{\sigma}_{n}(\widetilde{\omega},q^2,P^2)\equiv
\int \frac{d^4\xi}{\xi^4} e^{iq\cdot \xi} {\sigma}_{n}(P\cdot\xi,\xi^2,P^2),
\end{align}
where corresponding ${\cal O}_n$ can be any two-currents operator defined in Eq.~\eqref{eq:2currents}, and $\widetilde{\omega}\equiv \frac{2P\cdot q}{-q^2}=\frac{1}{x_B}$ with $x_B$ the Bjorken variable for the lepton-hadron deep inelastic scattering (DIS). Assuming $|P^2/q^2|\ll1$, $\widetilde{\sigma}_{n}$ has cuts going out to infinity from the thresholds $\widetilde{\omega}=\pm1$.  The reason for having the cuts is that the system has a positive energy when $\widetilde{\omega}^2>1$, corresponding to $(q+P)^2>0$ or $(q-P)^2>0$, and thus, it can produce physically propagating particles.
To understand the difference better, let us consider a simple integral which could contribute to Eq.~\eqref{eq:lcsM},
\begin{align}
\int \frac{d^4\xi}{\xi^4}\,\xi^{\nu}\,e^{i (q+ xP)\cdot \xi}\,,
\end{align}
where $-1<x<1$ could be thought as the momentum fraction of a parton inside of the hadron. If $q+ xP$ is not light-like for any value of $x$, which is equivalent to $\widetilde{\omega}^2<1$, this integral is always finite. But if $q+ xP$ is light-like, this integral is divergent in the region where $\xi$ is large and almost anti-collinear to $q+ xP$. This simple excise tells us that the non-analytic cut of $\widetilde{\sigma}_{n}$ comes from the integration region of large $\xi$. That is, even if we demand $|q^2|\gg \Lambda_{\text{QCD}}^2$, $\widetilde{\sigma}_{n}$ in momentum space can always receive contribution from large $\xi$ region so long as $\widetilde{\omega}^2>1$.  On the other hand,
in coordinate space, if we fix $\xi$ to be short-distance, we do not have contribution from the large $\xi$ region and thus ${\sigma}_{n}$ has a good analytic behavior.

Since $\widetilde{\sigma}_{n}$ is simply a Fourier transformation of ${\sigma}_{n}$, the factorization of ${\sigma}_{n}$ in Eq.~\eqref{eq:fac} implies the following factorization,
 \begin{align}\label{eq:facM}
\begin{split}
\widetilde{\sigma}_{n}=\sum_{a}\, f_{a} \otimes \widetilde{K}_{{n}}^{a} +O(\Lambda_{\text{QCD}}^2/q^2)\, ,
 \end{split}
\end{align}
where $\otimes$ represents the $x$-convolution in Eq.~(\ref{eq:fac}) and
\begin{align}\label{eq:Km}
\widetilde{K}_{n}^{a}=\int \frac{d^4\xi}{\xi^4}\,e^{i q\cdot \xi} K_{n}^{a}(x P\cdot\xi,\xi^2,x^2P^2,\mu)\,.
\end{align}
From the discussion above, the factorization in momentum space is unambiguous if $\widetilde{\omega}^2<1$.

\sect{Matching coefficients}
From Eq.~\eqref{eq:coeK}, we can obtain $K_{{n}}^a$ if we know the Wilson coefficients $W_{n}^{(J,a)}$. Our strategy to calculate them  is as follows: (1) calculate $K_{{n}}^a(x \omega,\xi^2,0,\mu)$, which corresponds to the coefficient function in collinear factorization with $P^2\to 0$, (2) expand $K_{{n}}^a(x \omega,\xi^2,0,\mu)$ as a power series of $x \omega$, and
(3) compare it with $K_{{n}}^a(x \omega,\xi^2,0,\mu)$ in Eq.~(\ref{eq:coeK}), along with the fact $\Sigma_J(x\omega,0) = (x\omega)^J$, to obtain $W_{n}^{(J,a)}$ as the expansion coefficients.

\begin{figure}[htb]
\begin{center}
\includegraphics[width=0.95\linewidth]{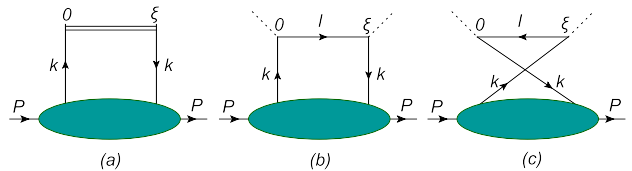}
\caption{\label{fig:LO}
LO Feynman diagrams for $\widetilde{\sigma}_n$.}
\end{center}
\end{figure}

In the following, we calculate nonvanishing $K_{{n}}^a$ for various LCSs introduced above to the lowest order (LO) in $\alpha_s$ expansion, which we denote as $K_{{n}}^{a(0)}$.
There is only one Feynman diagram shown in Fig.~\ref{fig:LO}(a) contributes to $K^{q(0)}_{q}$. According to our strategy above, we set $k^\mu=x P^\mu$ with $P^2=0$, and get
\begin{align}\label{eq:Kq0}
\begin{split}
K^{q(0)}_{q}(x \omega,\xi^2,0,\mu)&=\frac{1}{2} \text{Tr}[\slashed{k} \slashed{\xi}] e^{i\xi\cdot k}=2 x \omega  e^{i x \omega}\,,
\end{split}
\end{align}
which is consistent with the relation Eq.~\eqref{eq:rel2}. Two Feynman diagrams, as shown in Fig.~\ref{fig:LO}(b) and (c), contribute to $K^{q(0)}_{S}$ with
\begin{align}\label{eq:Mb}
M_b&=\frac{i\xi^4}{2}\int \frac{d^4 l}{(2\pi)^4}\frac{\text{Tr}[\slashed{k} \slashed{l}] e^{i\xi\cdot(k-l)}}{l^2+i\varepsilon}
=\frac{i}{\pi^2}\, x \omega\,  e^{i x \omega}\,,
\end{align}
and $M_c=M_b^*$.  We have the sum of these two diagrams,
\begin{align}\label{eq:Kqq0}
\begin{split}
K^{q(0)}_{S}(x \omega,\xi^2,0,\mu)&= \frac{i}{\pi^2}\, x \omega\left(
e^{i x \omega}-e^{-i x \omega}\right)\,,
\end{split}
\end{align}
which is consistent with the relation Eq.~\eqref{eq:rel1}. Results of $K^{q(0)}_{V}$ and $K^{q(0)}_{\widetilde V}$ are the same as $K^{q(0)}_{S}$ at this order. With $q\neq q'$, only Fig.~\ref{fig:LO}(b) contributes to $K^{q(0)}_{V^\prime}$, while only Fig.~\ref{fig:LO}(c) contributes to $K^{q^\prime(0)}_{V^\prime}$. Neglecting the mass of both $q$ and $q'$, we obtain
\begin{subequations}\label{eq:Vp20}
\begin{align}
K^{q(0)}_{V^\prime}(x \omega,\xi^2,0,\mu)&=\frac{i}{\pi^2}\, x \omega\,e^{i x \omega}\,,\\
K^{q^\prime(0)}_{V^\prime}(x \omega,\xi^2,0,\mu)&=\frac{-i}{\pi^2}\, x \omega\, e^{-i x \omega}\, .
\end{align}
\end{subequations}

Using our strategy and the results of $K_{{n}}^{a(0)}$ evaluated at ${P^2=0}$ above, we can easily derive $W_{n}^{(J,a)}$. For example,  by expanding $K^{q(0)}_{q}=\sum_{J=1} 2\,[i^{J-1}/(J-1)!] (x\omega)^J$, and comparing it with $K^{q(0)}_{q}=\sum_{J} 2\,W_q^{(J,q)}(x\omega)^J$ from Eq.~(\ref{eq:coeK}) at ${P^2=0}$, we obtain $W_{q}^{(J,q)}=i^{J-1}/(J-1)!$ with $J\geq 1$.  Substituting this into Eq.~(\ref{eq:coeK}),  we obtain $K^{q(0)}_{q}(x \omega,\xi^2,x^2P^2,\mu)$. For simplifying our discussion below, we assume $\xi^2$ small enough so that $P^2\xi^2$ (and thus $P^2$) terms can be ignored in the rest of this paper.

From Eq.~\eqref{eq:Km}, we can easily obtain $\widetilde{K}_n^a$ using the above calculated $K_n^a$. For example, we have
\begin{align}\label{eq:KmS0}
\widetilde{K}^{q(0)}_{S/V/\widetilde{V}}(x \widetilde\omega,q^2,0,\mu)&=-2i\,\frac{x^2\widetilde\omega^{2}}{1-x^2\widetilde\omega^{2}}\, ,
\end{align}
which has cuts going out to infinity from the thresholds $\widetilde\omega=\pm 1$ as discussed earlier.

\sect{Good ``lattice cross sections"}
After showing that the UV renormalized coordinate-space hadronic matrix elements ${\sigma}_{n}$ in Eq.~\eqref{eq:lcs} can be factorized to the PDFs with perturbatively calculable coefficient functions, we need to demonstrate that these matrix elements are calculable in LQCD with Euclidean time, in order for them to be good ``lattice cross sections''.  This can be easily satisfied if $\xi$ has no time component. In conclusion,  ${\sigma}_{n}$ defined in Eq.~\eqref{eq:lcs} are good LCSs for extracting PDFs under the condition  $\xi_0=0$. For example,  ${\sigma}_{S/V/\widetilde{V}}$ can be naturally used to extract $f_{q}(x,\mu)+f_{\bar q}(x,\mu)$, and ${\sigma}_{V^\prime}$ can be used to extract $f_{q}(x,\mu)+f_{{\bar q}^\prime}(x,\mu)$. With various linear combinations of $\sigma_n$, we could extract $f_{q}(x,\mu)$, $f_{\bar q}(x,\mu)$ and $f_{g}(x,\mu)$ individually.

If there are methods to calculate ${\sigma}_{n}$ or its linear combination in LQCD without setting $\xi_0=0$, then we can define more good LCSs accordingly. One possibility is that one can use Feynman-Hellmann technique to calculate $\widetilde{\sigma}_{n}$ with $q_0=0$~\cite{Orginos:2017kos} . Therefore, according to our above discussion, $\widetilde{\sigma}_{n}$ defined in Eq.~\eqref{eq:lcsM} are also good LCSs for extracting PDFs under the condition $q_0=0$ and $\widetilde{\omega}^2<1$.

Having many good LCSs makes it possible for extracting PDFs by using QCD global analysis of ``data'' generated by LQCD calculations of these LCSs. This will provide a promising way to determine PDFs from ab initio LQCD calculation. In fact, our method is so general that proposed LQCD calculable functions used in the literature to determine PDFs are special cases of these good LCSs, which we show in the following.

\sect{Relation to other methods}
Let us first discuss the relation to quasi-PDFs and pseudo-PDFs. With $K_q^{q(0)}$ calculated in Eq.~\eqref{eq:Kq0}, we can construct a linearly combined good LCS,
\begin{align}\label{eq:lcsQuasi}
\int \frac{d \omega}{\omega}\, \frac{e^{-i x \omega}}{4\pi} \sigma_{q}(\omega,\xi^2,P^2)\approx&f_{q}(x,\mu)\,,
\end{align}
modulo $O(\alpha_s)$ corrections and higher twist corrections. With $\xi_0=0$, the integral over $\omega=-\vec{\xi}\cdot\vec{P}=-|\vec{\xi}||\vec{P}|\cos\theta$ in Eq.~(\ref{eq:lcsQuasi}) could have different interpretations. Choosing both $\vec{P}$ and $\vec{\xi}$ along the ``3"-direction, which results in $\theta=0$, the left hand side of Eq.~\eqref{eq:lcsQuasi} is the quasi-quark distribution introduced in Ref.~\cite{Ji:2013dva} if the integral is performed by fixing $P_3$, while it is the pseudo-quark distribution used in Ref.~\cite{Orginos:2017kos}  if the integral is performed by fixing $\xi_3$. These two methods are equivalent if matching coefficients are calculated to lowest order in $\alpha_s$, but different if higher order contributions need to be considered. The Eq.~\eqref{eq:lcsQuasi} also tells us that, to effectively extract PDFs using good LCSs, one should generate lattice ``data" with as many different values of $\omega$ as possible.

From Eq.~\eqref{eq:KmS0} we have
\begin{align}\label{eq:facMS}
\begin{split}
\widetilde{\sigma}_{S/V/\widetilde{V}}&\approx -2i \widetilde{\omega}\int_{-1}^1 dx\,\frac{ x\widetilde{\omega}}{1-x^2\widetilde{\omega}^{2}} f_{q}(x,\mu^2)\, ,
 \end{split}
\end{align}
which is equivalent to $T_{33}$ used in Ref.~\cite{Chambers:2017dov} if we factorize the structure function therein to PDFs. More precisely, $T_{33}$ can be obtained if we construct momentum-space good LCSs $\widetilde{\sigma}_{\mu\nu}$ using operator ${\cal O}_{\mu\nu}$ defined in Eq.~\eqref{eq:operatorsT}, and then set $\mu=\nu=3$ and $P_3=q_3=q_0=0$. However, because $\widetilde{\sigma}_{\mu\nu}$ has effective only two degrees of freedom, all of its nontrivial information should have been encoded in $\widetilde{\sigma}_{V}$ and $\widetilde{\sigma}_{\widetilde V}$. We thus expect that extracting PDFs using  $\widetilde{\sigma}_{V}$ and $\widetilde{\sigma}_{\widetilde V}$ can be at least as good as that using $T_{33}$.

\sect{Summary}
We constructed a class of good ``lattice cross sections" in terms of single-hadron matrix elements of renormalized equal-time nonlocal operators in coordinate space, as defined in Eq.~\eqref{eq:lcs}.  We show that these matrix elements are calculable in LQCD and factorizable to PDFs with perturbative coefficients, so long as the largest separation between fields of the operator satisfies $|\vec{\xi}|\ll1/\Lambda_{\text{QCD}}$.  We also show that corresponding momentum-space matrix elements, $\widetilde{\sigma}_n$ defined in Eq.~\eqref{eq:lcsM} with $|\frac{2P\cdot q}{-q^2}|<1$ and $|q^2|\gg\Lambda_{\text{QCD}}^2$, are also good LCSs.

We provided some explicit examples of good LCSs made of quark fields in Eqs.~\eqref{eq:operators} and \eqref{eq:operators2}, calculated corresponding matching coefficients to LO in $\alpha_s$, and connect the matrix elements to unpolarized PDFs. As briefly mentioned earlier, it is straight forward to construct the gluonic operators to have good LCSs directly sensitive to gluon distributions.  Our approach could be easily applied for polarized PDFs and other parton correlation functions of various hadrons.  We also find that some LQCD calculable functions used in literature for extracting PDFs, including quasi-PDFs proposed in Ref.~\cite{Ji:2013dva}, pseudo-PDFs used in Ref.~ \cite{Orginos:2017kos} and $T_{33}$ used in Ref.~\cite{Chambers:2017dov}, are special cases of good LCSs.  With the large number of identified good LCSs, we could extract PDFs and hadron structure in general by QCD global analyses of the data generated by ab initio LQCD calculations of these LCSs.

Since it is very easy to generate lattice data with small $\omega$, LQCD calculations could provide the much needed information on PDFs at relatively large $x$, complementary to what could be extracted from experimental measurements. In addition, LQCD calculations could provide the information on partonic structure of hadrons, such as neutron, pion or keon, that could be difficult to do scattering experiment with.

\sect{Acknowledgments}
We thank Chuan Liu, Xiaohui Liu, A.H.~Mueller, and Feng Yuan for helpful communications. Especially, we thank Xu Feng for plenty of useful discussions. This work is supported in part by the U.S. Department of Energy, Office of Science, Office of Nuclear Physics under Award No.~DE-AC05-06OR23177, within the framework of the TMD Topical Collaboration.


\end{document}